\documentclass[aps,prl,floatfix,twocolumn,showpacs,superscriptaddress]{revtex4}
\usepackage{graphicx}
\usepackage{amssymb}
\usepackage{amsmath}
\usepackage{soul}
\usepackage[usenames]{color}
\usepackage{ulem}

\ifx\pdftexversion\undefined
\usepackage[dvips]{hyperref}
\else
\usepackage{hyperref}
\fi
\hypersetup{
  colorlinks = true, linkcolor = blue
}

\renewcommand{\phi}{\varphi}
\renewcommand{\theta}{\vartheta}

\def \be{\begin{equation}}
\def \ee{\end{equation}}
\def \ba{\begin{array}}
\def \ea{\end{array}}
\def \bea{\begin{eqnarray}}
\def \eea{\end{eqnarray}}

\def \half{\frac{1}{2}}

\def \w{{\omega}}

\def \nd{{^{\vphantom{\dagger}}}}
\def \yd{^\dagger}

\begin{document}

\title{Quasiparticle dynamics in a Bose insulator probed by inter-band Bragg spectroscopy}
\author{N.~Fabbri}
\email[Electronic address: ]{fabbri@lens.unifi.it}
\address{LENS, Dipartimento di Fisica e Astronomia, Universit\`a di Firenze and INO-CNR, via Nello Carrara 1, I-50019 Sesto Fiorentino (FI), Italy}

\author{S.~D.~Huber}
\email[Electronic address: ]{sebastian.huber@weizmann.ac.il}
\address{Department of Condensed Matter Physics, The Weizmann Institute of Science, Rehovot, 76100, Israel}

\author{D.~Cl\'ement}
\altaffiliation[Present address: ]{Laboratoire Charles Fabry, 2 avenue Agustin Fresnel, 91127 Palaiseau Cedex, France}
\address{LENS, Dipartimento di Fisica e Astronomia, Universit\`a di Firenze and INO-CNR, via Nello Carrara 1, I-50019 Sesto Fiorentino (FI), Italy}

\author{L.~Fallani}
\affiliation{LENS, Dipartimento di Fisica e Astronomia, Universit\`a di Firenze and INO-CNR, via Nello Carrara 1, I-50019 Sesto Fiorentino (FI), Italy}

\author{C.~Fort}
\address{LENS, Dipartimento di Fisica e Astronomia, Universit\`a di Firenze and INO-CNR, via Nello Carrara 1, I-50019 Sesto Fiorentino (FI), Italy}

\author{ M.~Inguscio}
\address{LENS, Dipartimento di Fisica e Astronomia, Universit\`a di Firenze and INO-CNR, via Nello Carrara 1, I-50019 Sesto Fiorentino (FI), Italy}

\author{E.~Altman}
\address{Department of Condensed Matter Physics, The Weizmann Institute of Science, Rehovot, 76100, Israel}

\begin{abstract}
We investigate experimentally and theoretically the dynamical properties of
a Mott insulator in decoupled one-dimensional chains. Using a theoretical
analysis of the Bragg excitation scheme we show that the spectrum of
inter-band transitions holds information on the single-particle Green's
function of the insulator. In particular the existence of particle-hole
coherence due to quantum fluctuations in the Mott state is clearly seen in
the Bragg spectra and quantified. Finally we propose a scheme to directly
measure the full, momentum resolved spectral function as obtained in
angle-resolved photoemission spectroscopy of solids.
\end{abstract}


\maketitle

The observation of the superfluid (SF) to Mott insulator (MI) transition of bosons in optical lattices \cite{Greiner02}
has received considerable attention as a paradigmatic example of a quantum phase-transition driven by interactions. The properties of lattice bosons in this strongly correlated regime have been probed using several methods
\cite{Stoferle04, Foelling05, Campbell06, Spielman08, Du10, Clement09, Bakr10, Sherson10,Bissbort11}.  For example, time-of-flight experiments were used to study the development of spatial first-order coherence over increasing
length-scales inside the Mott state upon approaching the transition to the SF
state \cite{Foelling05}. In a quantum system, the emergence of such spatial
correlations must go hand in hand with increasing temporal correlations.
Near the quantum critical point the precise relation between the two is
determined by the dynamical critical exponent of the transition (see {\it
e.g.} Ref.~\cite{Sachdev99}).  Away from the transition, where critical
properties are not yet apparent, the temporal first-order coherence lends
insight on the nature of the quasi-particle excitations of the strongly
correlated state.

So far however, most of the dynamical experiments in the Mott regime
using schemes of lattice modulation and Bragg spectroscopy have focused on
excitation frequencies matching transition within the lowest-energy Bloch
band \cite{Stoferle04, Clement09}. In this case, the external perturbation
is coupled to density fluctuations and in the linear response regime
the absorption spectrum is directly related to the dynamical structure
factor $S(q, \omega)$ of collective excitations, or particle-hole spectra
\cite{Iucci05,Rey2005,Huber07}.

\begin{figure}[ht!]
\begin{center}
\includegraphics[width=0.8\columnwidth]{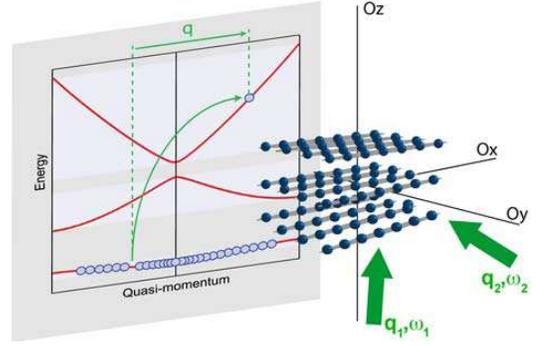}
\end{center}
\caption{
(color online) An array of 1D gases created by a 2D optical lattice
is driven into the MI by a third OL in the $Ox$ direction. The energy
band structure in the 1D lattice is depicted on the left with red solid
lines. The lowest band corresponds to particles in the MI and above it
are the higher single-particle energy bands. Two laser beams (Bragg beams,
green) excite the initial MI by transferring a particle to a high-energy
band and leaving a hole in the many-body state.
}
\label{fig:scheme}
\end{figure}

In this Letter, we show how inter-band Bragg spectroscopy
\cite{Clement09,Heinze11} supplemented by a theoretical model of the Mott
insulator can be used to extract properties of single-particle excitations
in the many-body state. We then suggest a refined approach for directly
measuring the single-particle Green's function in a model-independent way.

The MI state is realized in decoupled one-dimensional (1D) chains of
interacting bosons, as represented in Fig.~\ref{fig:scheme}.
 We excite the system with two simultaneous laser pulses (Bragg beams) which
 induce an energy transfer $\hbar \omega= \hbar (\omega_{1}-\omega_{2})$
($\omega_{1,2}$ being the laser  beams frequencies)
and a momentum transfer $\hbar {\bf q}=\hbar {\bf q}_{1}-\hbar{\bf q}_{2}=\hbar
q {\bf e}_{x}$ along the axis of the 1D  chains (${\bf q}_{1}$ and
${\bf q}_{2}$ being the wave-vectors of the Bragg photons). We measure
the energy absorption spectrum $D(\omega)$ at a fixed momentum transfer
$q\lesssim\pi/a$, where $a$ is the periodicity of the lattice along the
chains. We show how, with the precise knowledge of the particle dispersion in the
high band \cite{Abramowitz65-Mathieu,Zwerger03}, it is possible to obtain
information about quasi-particle structure and dynamics in the MI. Moreover,
a refined scheme would give access to a momentum resolved absorption
rate $D(k,\w)$, which we show is directly related to the single-particle
spectral function $A(k,\w)$ in the lowest-energy band. From the latter,
one obtains the Green's function (GF) of the MI, lending information on
both spatial and temporal coherence of quasiparticles. Such a measurement is
analogous to angle-resolved photoemission spectroscopy used in solid-state
\cite{Damascelli03}, recently extended to ultracold gases
through Raman \cite{Dao07,Dao09} and rf spectroscopy \cite{Stewart2008}.

We start by  briefly describing the experimental setup used to obtain the inter band spectra (details can be found in \cite{Clement09,Clement09a}). We load a Bose-Einstein
condensate of $^{87}$Rb in a 3D optical lattice at the wavelength
$\lambda_{L}=830~$nm. The amplitudes $V_{i}$ of the lattices along each axis
$i=x,y,z$ are expressed in units of the recoil energy $E_{\rm R}=h^2/(2 m
\lambda_{L}^2)$, $V_{i}=s_{i}E_{\rm R}$, where $m$ denotes the mass
of $^{87}$Rb. The optical lattices are ramped up to their final values
$s_{i}$ with an exponential ramp of duration $140\,$ms and time constant
$30\,$ms. Two lattice amplitudes ($s_{y}\!=\!s_{z}\!=\!35$) are fixed to
create an array of 1D  chains. The amplitude of the third lattice $s_{x}$
is varied to tune the ratio between the on-site interaction energy $U$
and the tunneling amplitude $J_{1}$ between Wannier states of the lowest Bloch band
in each 1D MI chain from
$U/2J_{1} \simeq 7$ to $U/2J_{1} \simeq 42$.

The Bragg beams are derived from a laser at $780\,$nm, detuned by $\sim
200\,$GHz from the D$_{2}$ line of $^{87}$Rb. In this work we fix $q=0.96
\pi/a$ and we measure the amount of excitations  induced by the Bragg beams
as a function of their frequency difference $\omega$. The measured quantity
$D(\omega)$ is the mean square width of the zero-momentum peak in a phase-coherent lattice-gas obtained after lowering the 3D optical lattices. Timing and details of the experimental procedure can be found in \cite{Clement09a}, in particular the way $D(\omega)$ is re-scaled with the parameters of the Bragg beams to allow a relative comparison of the different spectra. In \cite{Clement09a}, we also verified that $D(\omega)$ is proportional to the energy transferred to the gas so that it  can be written as $D(\omega)= \mathcal{C} \ \omega S(q,\omega)$  where $\mathcal{C}$ is a constant independent of the lattice strength \cite{Brunello01}.

\begin{figure}[t]
\begin{center}
\includegraphics{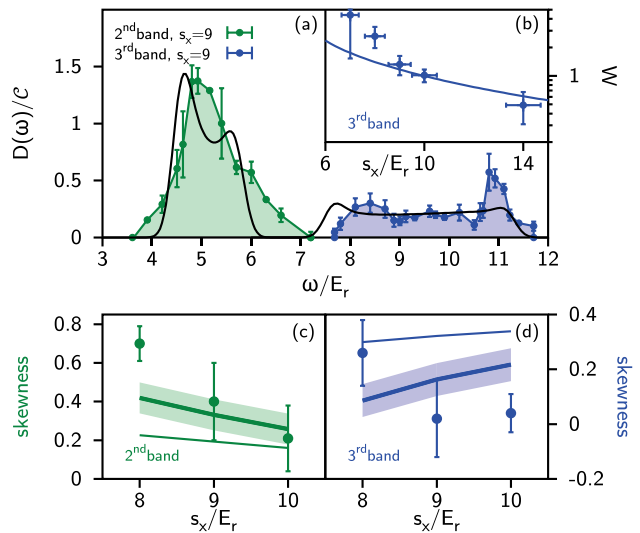}
\end{center}
\caption{
(color online) {\bf Inter-band Bragg spectra.}(a) Energy
absorption rate $D(\omega)$ over the energy range of the second and third
Bloch band for a Bragg momentum transfer $q=0.96\pi/a$ and for a lattice
depth $s_{x}=9$. The (blue and green) dots are the experimental data with
error bars indicating statistical uncertainties after averaging over 4 to
5 experimental acquisitions, the black lines are the theoretical predictions
for $D(\omega)$. (b) Integrated spectral weight $W$ in the third band.  (c) and (d) Skewness (third moment) of $S(q,\w)$ in
the second (c) and third (d) band. The measured values are compared to the theoretical prediction  including particle-hole coherence (thick lines) and to the contribution of the single-particle density of states alone (thin lines). The
shaded area indicates the systematic uncertainty in determination of $s_x$ in the experiment.
}
\label{fig:lineshape}
\end{figure}

In Fig.~\ref{fig:lineshape}(a) we show an example of the spectra
obtained for lattice of amplitude $s_x\!=\!9$ at frequencies resonant with
transitions to the second and  third Bloch bands. The total spectral weight $W\!=\!\int\! d\omega D(\omega)$  of  transitions to the third band as a function of $s_x$ is shown in
Fig.~\ref{fig:lineshape}(b). The suppression of spectral weight is due to reduction of the matrix element, or Frank-Condon overlap, between wavefunctions of the two bands with increasing lattice strength. The distribution of spectral weight within each band is primarily determined by two factors (i) the density of final states (DOS) in that band and (ii) the particle hole coherence in the Mott ground state driven by quantum fluctuations about the classical state with precisely $n$ (integer) particles on each site.
The peaks seen at the band edges, \textit{e.g.}, are the result of the divergent DOS there. Furthermore we quantify the asymmetry of the spectra about the band centers through the skewness (third moment) \cite{skew} of $S(q,\w)$. This is shown in Fig.~\ref{fig:lineshape}(c) and (d) for the second and third band as a function of $s_x$. A positive skewness corresponds to an imbalance towards lower energies. Using a theoretical model we shall relate the skewness and its variation with band index and lattice amplitude, to the particle-hole coherence in the Mott state.

The Bragg perturbation couples to the particle-density and, in the linear response regime, the excitation rate $D(\omega)$ is
directly related to the dynamic structure factor
\begin{equation}
\label{eqn:sqw}
S(q,\omega)=
\sum_{m}|\langle m|\rho_{q}|0\rangle|^{2}\delta(\omega-\omega_{m0})=
{\rm Im}[\Pi(q,\w)],
\end{equation}
where $\rho_{q}$ is the density operator at momentum $q$
\cite{Stenger99}. The sum runs over excited states and $\hbar \omega_{m0}$
is the associated excitation energy.

\begin{figure}[tb]
\begin{center}
\includegraphics[width=0.7\columnwidth]{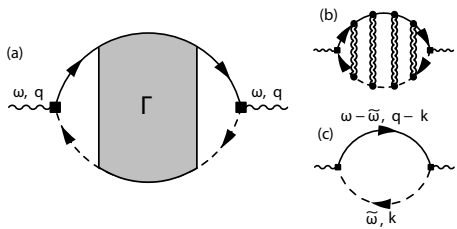}
\end{center}
\caption{
{\bf Response functions.} (a) Diagram describing the Bragg excitation in
the linear response. The black squares describe the coupling of the Bragg
beams to the density fluctuations. The full (dashed) lines denote the GF of
a hole in the MI (of an upper band particle). The gray area $\Gamma$ stands
for the final-state interactions between the upper band particle and the MI
hole. (b) In a $T$-matrix approximation the final state interaction gives
rise to a ladder diagram where the wiggly lines denote the interaction
between upper band particle and the hole. (c) In the absence of final
state interactions the bare bubble describes the experimental response.
}
\label{fig:diagrams}
\end{figure}

The response function $\Pi(q,\w)$ is represented graphically by the bubble
diagram in Fig \ref{fig:diagrams}(a): (i) the vertex de\-scri\-bes the coupling of the Bragg beams to the density operator $\rho_q$; (ii) the full line is the GF of the hole produced in the lowest band by the Bragg excitation; (iii) the dashed line represents the GF of the particle excited to the $n$th band; (iv) the filled area is the $T$-matrix for scattering of the
upper-band particle with the hole in the lowest band. We explain how we evaluate each part of the diagram, showing that for excitations
to the higher bands the contribution of the final-state interaction can be neglected, hence the process can be well described by the bare bubble shown in Fig. \ref{fig:diagrams}(c). Since without this interaction the
propagation of the upper-band particle can be calculated as the one of a free particle, the experiment effectively probes the remaining unknown which is related to the single-particle GF.

Let us first consider the edge vertices which correspond to the matrix element for the excitation of a particle from the lowest to the upper band by the density operator. To find this matrix element we express the field operators in terms of creation operators in Wannier states of the lattice sites, so that $\rho(x)=\sum_{ij,nm}
w^\ast_{mj}(x)w_{ni}(x)b\yd_{ni}b\nd_{mj}$, where $n,m$ are band indices,
$i,j$ site indices and $w_{ni}(x)$ the respective Wannier functions. From now on we assume that the lattice is sufficiently
deep that the overlap between Wannier functions of neighboring sites can be
neglected in the density operator. In addition, we focus on the component
of the density operator that induces transitions from the lowest to the
$n$th band:
\begin{equation}
\label{eqn:density}
\rho_{n}(q)=
\bar\rho F_{1n}(q)\sum_{i} e^{-iqR_{i}} b_{ni}^{\dag}b_{1i},
\end{equation}
where $R_i$ is the position at site $i$ and $\bar\rho$ the filling of the MI.
$F_{1n}(q)$ is the matrix element
\begin{equation}
F_{1n}(q)\!=\! \int \! dx\, w_{1i}(x)e^{-iqx}w_{ni}(x)
\approx \frac{(-i q l_0)^{n}}{\sqrt{2^{n}n!}}e^{-\frac{(ql_{0})^{2}}{4}}.
\end{equation}
To get a simple expression for the dependence of $F_{1n}(q)$ on the lattice strength we approximated the Wannier states by those of a harmonic well with oscillator length $l_{0}=\lambda/(2\pi s_{x}^{1/4})$; however for the comparison with experimental results we use the exact Wannier functions \cite{Zwerger03}.

To describe the hole propagation in the MI we use the generalized Bogoliubov
theory \cite{Altman02, Huber07}, which accounts for quadratic quantum
fluctuations (i.e. virtual particle and holes) about the classical (Gutzwiller) ground-state.
The particle operator in the lowest band
can be represented in terms of the Bogoliubov quasi-particle and quasi-hole excitations as $b\nd_k=\sqrt{f(k)}(\beta\yd_{h,k}+\beta_{p,k})$, which are a combination of an added particle (doublon) and removed hole on the Mott background, \textit{e.g.}, $\beta^{\dagger}_{p,k} = u_k p^{\dagger}_k + v_k h_{-k}$ and $f(k) = (u_k - v_k)^2$. The ground state is the vacuum of the operators $\beta_{p/h,k}$. Therefore $f(k) =\langle b^{\dagger}_k b_k\rangle$ is simply the momentum distribution.
Accordingly the single particle Green's function in the lowest Bloch band is given by:
\be
G_1(k,i\w)={f(k)\over i\w-\w_p(k)}+ {f(k)\over -i\w-\w_h(k)}\equiv G_p+G_h.
\ee
$\omega_p(k)$ (\textit{resp.} $\omega_{h}(k)$) denotes the dispersion
relation of a particle (\textit{resp.} a hole) in the lowest Bloch band.
The momentum distribution $f(k)$ stems from quasiparticle coherence factors
and within the Bogoliubov theory it is
\be
f(k)= {1\over \sqrt{1-{J_1 \over J_c}\cos k}} .
\ee
Here $J_c$ is the critical hopping strength at the Mott transition.
We shall restrict ourselves to using the Bogoliubov theory deep in the Mott insulator
where it provides a good approximation of the single particle
Green's function \cite{note-conserving}. In this regime $f(k)\approx 1+\half(J_1/  J_c)\cos k$ to leading order in $J_1/J_c$.

The GF of a single particle in the $n$th upper band is taken to be that of a
free particle with appropriate band dispersion $\tilde \omega_{n} (k)$. We
take into account a slightly renormalized dispersion due to interaction
with the background of filled sites of the MI \cite{supplementary}.

The interaction between the particle in the upper band and the hole
in the lowest band is included in the full $T$-matrix (filled box in
Fig. \ref{fig:diagrams}(a)). In general this leads to a complicated sum of
diagrams including all possible sequences of multiple collisions through
the interaction term $U_{1n} b\yd_{ni}b\nd_{ni} (p\yd_{i}p\nd_i-h\yd_i
h\nd_i)$. Here we represented the interaction in terms of actual particles
and holes in the classical ground state, where $U_{1n}$ is the interaction
matrix element between Wannier states of the lowest and the $n$th band. The
interaction looks more complicated when expressed in terms of the Bogoliubov
quasi-particles and quasi-holes. However, the sum simplifies in the strong
lattice limit when $v_k\ll u_k$. Then, to leading order in $v_k$ we can
include only the ladder diagrams shown in Fig. \ref{fig:diagrams}(b),
which are easily summed up as a geometric series \cite{supplementary}. The
result of the interaction, treated by the ladder summation, is to induce
a bound state between the upper band particle and the hole in the MI
\cite{supplementary}. For the experimental parameters the weight carried
by this bound state ($<1\%$) is too small to affect the measurements. We
conclude that to an excellent approximation we can use the bare bubble
diagram shown in Fig. \ref{fig:diagrams}(c) to compute the structure factor.

The structure factor computed from the bare bubble diagram is given by
\begin{equation}
\label{eqn:resultsqw}
\!\!S(q,\omega)=\bar\rho^{2}|F_{1n}|^{2}\int\frac{dk}{2\pi},
A_h(k-q,\w-\tilde{\w}_n(k))
\end{equation}
where $A_h(k,\w)=-\pi^{-1}\text{Im}\, G_h(k,\w+i0^+)$ is the hole spectral function in the Mott insulator. In particular, the spectral function obtained from the generalized Bogoliubov theory of the Mott insulator is
$A_h(k,\w)=f(k)\delta[\omega-\omega_{h}(k)]$.

This is the formula we use to compare with the experiment. However to clearly reveal the two important factors in the spectra it is worth making another simplification. In the regime of interest, of strong optical lattice, the bandwidth of the hole in the lowest band can be neglected compared to the dispersion of a particle in the excited band and we can take $\w_h(k)=\w_0 $. This leads to
\be
S(q,\w)\approx \bar\rho^{2}|F_{1n}(q)|^{2} \rho_n(\w-\w_0) f(k_n(\w-\w_0)-q),
\label{eqn:sqwapprox}
\ee
$\rho_n(\w)$ being the single particle DOS in the $n$th band and $k_n(\w)$ the inverse function of the dispersion in that band $\w_n(k)$.

In the limit of infinitely deep lattice $f(k)\to 1$ and the observed line-shape is determined solely by the single particle DOS. With reduction of the lattice amplitude (increased hopping) $f(k)$
becomes more strongly peaked near $k=0$ and therefore contributes more
significantly to the line-shape. Specifically it gives increased weight to frequencies resonant with transitions that  create a Mott hole near $k=0$ and an excited $n$th band particle with quasi-momentum $q$.
If we take $q\approx \pi/a$, as in the experiment, this effect enhances the weight of transitions that create a higher band particle near the Brillouine zone edge. From the band structure shown in Fig. \ref{fig:scheme} it is clear that in this way the momentum distribution skews the spectra of the second band toward lower energies and those of the third band toward higher energies. It should be noted however that the single-particle DOS is itself not symmetric about the band centers. In particular for both the second and third band the peak of the DOS is skewed toward lower energies (positive skewness). Therefore, the effect of coherence in the Mott insulator is to increase the skewness of the second band and decrease it in the third band spectra.

Using Eq.~(\ref{eqn:resultsqw}) we calculate the experimental observable
$D(\omega)=\mathcal{C} \omega S(q,\omega)$. The proportionality constant $\mathcal{C}$ is fixed by matching the integrated spectral weight of excitations to the third band $W$ for a single
value of the lattice strength ($s_x=10$). We use the same constant to compute the spectra for all other lattice amplitudes and for all the bands. In addition, we broaden the delta-function
in Eq.~(\ref{eqn:resultsqw}) to effectively account
for the trap confining potential.

The spectrum obtained in this way  is presented and compared to the experimental results
in Fig.~\ref{fig:lineshape}(a) for $s_x=9$. Note that there are no free parameters except the overall proportionality constant $C$ which was calibrated once. We attribute the relative shift of the spectra to the systematic uncertainty in the actual lattice amplitude in the experiment\cite{note}. The calculated total weight of absorption $W$, shown
in Fig.~\ref{fig:lineshape}(b), is in good agreement with the experimental data. The reduction of $W$
with increasing lattice strength is due to suppression of the Frank-Condon factor $F_{1n}$.

As discussed above, the skewness of the structure factor relative to that of the
pure single particle density of states is a direct measure of the quasi-particle coherence factor in the Mott insulator. Fig. ~\ref{fig:lineshape} (c) and (d) compares the measured skewness  \cite{skew} of $S(q,\w)$ in the second and third bands to that calculated from the theoretical spectra (thick lines). Both are compared to the skewness of the single particle DOS in the corresponding bands. As anticipated, the actual skewness is consistently higher than the pure DOS effect in the second band and lower than the DOS effect in the third band and this effect is observed in a systematic way for different lattice amplitude. This is a clear indication of coherence effects inside the MI.

To conclude, we have shown that the inter-band Bragg absorption spectrum, in the linear response regime, gives information on particle Green's function. In particular we have quantified the single hole coherence through analysis of the asymmetry of the spectra.

It is important to note that the structure is related, through Eq. (\ref{eqn:resultsqw}) to a rather complicated weighted sum over the hole spectral function and is not proportional to the spectral function itself. For this reason input from a theoretical model of the Green's function in the Mott insulator was needed to extract a measure of the hole coherence. For this we used a Bogoliubov-like theory that takes into account zero-point fluctuations around the mean field Mott wave function. The non-trivial hole coherence stems from these zero point fluctuations.

It would be interesting to directly measure the single hole spectral function and thereby obtain quasi-particle energies, coherence factors and decay times in a model independent way. We suggest that in principle
this can be done using a band mapping technique \cite{Greiner01}. By counting how many particles are excited to a particular $k$ state in the upper band we would eliminate the $k$ integral in Eq. (\ref{eqn:resultsqw}). Then the response function corresponding to the excitation rate per final momentum $k$ would  be
\be
 S_{1n}(q,k,\w)=\bar{\rho}^2 |F_{1n}(q)|^2 A_h(k-q,w-\tilde\omega_n(k)).
\ee
This theoretical description of the measurement process is
identical to the description of angle-resolved photoemission spectroscopy
\cite{Caroli73,Mahan90}, a method that is currently used extensively
to measure the electronic spectral function of interesting materials
\cite{Damascelli03}. Our proposed scheme could also be implemented on more
complex many-body ground-states in a lattice.

This work was supported by ISF, MAE-MIUR (Joint
Laboratory LENS-Weizmann), ECR Firenze, by ERC through the DISQUA grant, and by MIUR through PRIN2009. S.D.H acknowledges support
by the Swiss Society of Friends of the Weizmann Institute of Science.


\begin{thebibliography}{21}
\bibitem{skew} The skewness is defined as $\gamma=\frac{1}{W\sigma^3} \int
d\omega (\omega-\bar\omega)^3 S(q,\omega)$, where $\sigma^2=\frac{1}{W}\int
d\omega (\omega-\bar\omega)^2 S(q,\omega)$ and $\bar \omega = \frac{1}{W}
\int d\omega \omega S(q,\omega)$.
\bibitem{note-conserving} On approaching the transition the Bogoliubov expression of the single particle Green's function suffers from being a non-conserving approximation. The
    expression for density response function does not suffer from the same problem \cite{Huber07}.
\bibitem{note}  Note that a relative frequency shift between the experimental
data points and the theoretical curves can be explained as a result of
the 10\% systematic uncertainty in the lattice height calibration. As
a matter of fact the position of the third Bloch band shifts by 0.3Er
(0.4Er) passing from s=8 to s=8.8 (s=10 to s=11).
\end{thebibliography}
\end{document}


\title{Supplementary material for the manuscript ``Quasiparticle dynamics in a Bose insulator probed by inter-band Bragg spectroscopy''}
\author{N.~Fabbri}
\email[Electronic address: ]{fabbri@lens.unifi.it}
\affiliation{LENS, Dipartimento di Fisica e Astronomia, Universit\`a di Firenze and INO-CNR, via Nello Carrara 1, I-50019 Sesto Fiorentino (FI), Italy}
\author{S.~D.~Huber}
\email[Electronic address: ]{sebastian.huber@weizmann.ac.il}
\affiliation{Department of Condensed Matter Physics, The Weizmann Institute of Science, Rehovot, 76100, Israel}
\author{D.~Cl\'ement}
\altaffiliation[Present address: ]{Laboratoire Charles Fabry, Institut d'Optique Graduate School, 2 avenue Agustin Fresnel, 91127 Palaiseau Cedex, France}
\affiliation{LENS, Dipartimento di Fisica e Astronomia, Universit\`a di Firenze and INO-CNR, via Nello Carrara 1, I-50019 Sesto Fiorentino (FI), Italy}
\author{L.~Fallani}
\affiliation{LENS, Dipartimento di Fisica e Astronomia, Universit\`a di Firenze and INO-CNR, via Nello Carrara 1, I-50019 Sesto Fiorentino (FI), Italy}
\author{C.~Fort}
\affiliation{LENS, Dipartimento di Fisica e Astronomia, Universit\`a di Firenze and INO-CNR, via Nello Carrara 1, I-50019 Sesto Fiorentino (FI), Italy}
\author{E.~Altman}
\affiliation{Department of Condensed Matter Physics, The Weizmann Institute of Science, Rehovot, 76100, Israel}
\author{ M.~Inguscio}
\affiliation{LENS, Dipartimento di Fisica e Astronomia, Universit\`a di Firenze and INO-CNR, via Nello Carrara 1, I-50019 Sesto Fiorentino (FI), Italy}

\maketitle

\section{Mean-field theory for the Mott insulator}

The Mott insulator which is realized in the deep lattice limit $U\gg J_{1}$ can be described using a strong-coupling mean-field approach \cite{Huber07}. In spirit it
is similar to the Bogoliubov theory for weakly interacting particles
\begin{align}
\nonumber
\beta_{p,k}^{\dag} &= \phantom{-}u_{k} p_{k}^{\dag} + v_{k} h_{-k}, \\
\nonumber
\beta_{h,k}^{\dag} &= -v_{k} p_{k} - u_{k} h_{-k}^{\dag}, \\
\label{eqn:meanfieldstatessupp}
u_{k} &=
\cosh
\left\{
\frac{1}{2}
{\rm atanh}\left[\frac{\frac{J_{1}}{J_{c}}\cos(k)}
{2-\frac{J_{1}}{J_{c}}\cos(k)}\right]
\right
\},
\\
\nonumber
v_{k} &=
\sinh
\left\{
\frac{1}{2}
{\rm atanh}\left[\frac{\frac{J_{1}}{J_{c}}\cos(k)}
{2-\frac{J_{1}}{J_{c}}\cos(k)}\right]
\right
\},
\\
\nonumber
b_{1,i}^{\dag} &\approx p_{i}^{\dag} + h_{i},
\;\; \mbox{and} \;\;
b_{1,i}^{\dag}b_{1,i} =
 \bar\rho+p_{i}^{\dag}p_{i}\!-\!h_{i}^{\dag}h_{i}.
\end{align}
Here, $\beta_{p/h,k}^{\dag}$ create a quasi-particle and a quasi-hole
excitation, respectively. The parameters $u_{k}$, $v_{k}$ behave as
$u_{k}\rightarrow 1$ and $v_{k} \rightarrow 0$ for $J_{1}/U\rightarrow
0$ and the combination $u_{k}v_{k}$ reflects the coherence in the Mott
insulator. The operators $p_{i}^{\dag}$ ($h_{i}^{\dag}$) create a state with
an additional (missing) particle at site $i$ on top of the classical
Gutzwiller state $\propto\prod_{i} (b_{1,i}^{\dag})^{\bar\rho} | {\rm
vac}\rangle$. Within a mean-field approach the effective Hamiltonian for
the Mott insulator reads
\begin{align}
H_{\ssm MI}&=\sum_{k} \omega_{h}(k)
\beta_{h,k}^{\dag}\beta_{h,k}+\omega_{p}(k) \beta_{p,k}^{\dag}\beta_{p,k}, \\
\omega_{p/h}(k)&=\frac{U}{2}\sqrt{1-(J_{1}/J_{c})\cos(ka)}\pm \mu,
\end{align}
where $J_{c}\approx 8U\bar\rho$ is the critical hopping for the formation
of the Mott state with filling $\bar\rho$ and $\mu$ denotes the chemical
potential. Note, that since the gap cuts off the infra-red divergencies characteristic of gapless 1d systems, the mean field theory is expected to work well inside the Mott phase. Details
about the mean-field approach can be found in Ref.~\cite{Huber07}.

\section{Final state interactions}

Let us now turn to the higher Bloch bands and the interaction effects
therein. We treat the interactions by splitting the field operators
into Wannier states for the lowest band and keep continuum states for
the higher bands $\psi^{\dag}(x) = \sum_{i}w_{1i}(x)b^{\dag}_{1i} +
\tilde\psi^{\dag}(x)$. The part of the interaction which describes scattering between the final state and the remaining particles in the lower band is:
\begin{equation}
\label{eqn:interbandint}
H_{\ssm int} = \frac{2\pi a_{s}\hbar^{2}}{m} \int dx\sum_{i} \,
w_{1,i}^{2}(x)  b_{1i}^{\dag} b_{1i}
\tilde \psi^{\dag}(x) \tilde \psi(x).
\end{equation}

For later convenience let us also express the density operator in the lowest band in terms of the particle and hole fluctuation operators:
\begin{align}
\nonumber
b_{1i}^{\dag}b_{1i} &= \langle b_{1i}^{\dag}b_{1i}\rangle +
(b_{1i}^{\dag}b_{1i}- \langle b_{1i}^{\dag}b_{1i}\rangle)\\ &=
\bar\rho + p_{i}^{\dag}p_{i}-h_{i}^{\dag}h_{i}.
\label{eqn:defbeta}
\end{align}

\begin{figure}[tb]
\begin{center}
\includegraphics{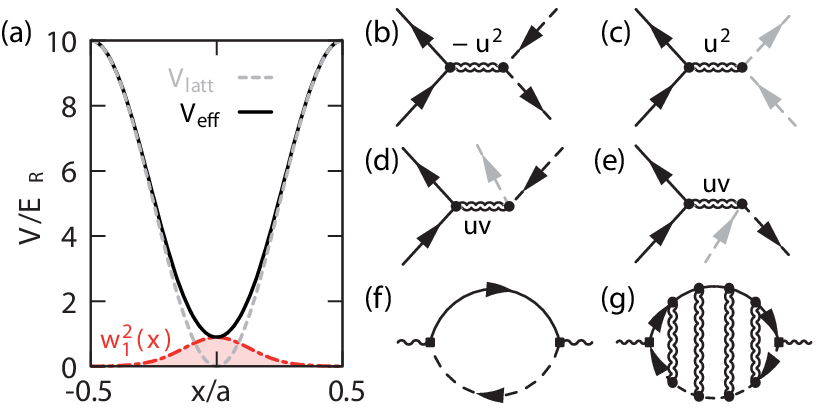}
\end{center}
\caption{
(color online) {\bf Final state interactions.} (a) The effective potential
for higher Bloch bands due to the underlying Mott insulator. (b)--(e)
Scattering processes due to the inter-band interactions. A full line
denotes a particle in the excited Bloch band, a black (gray) dashed line a
quasi-hole (quasi-particle) in the Mott insulator. Note that quasi-holes
and excited-band particles attract each other. (f) Polarization bubble
corresponding to $S_{n}(q,\omega)$. (g) Interaction corrections to
$S_{n}(q,\omega)$ leading to a bound state, a Mahan-exciton.
}
\label{fig:interactions}
\end{figure}

\paragraph{Hartree shift.} The Hartree approximation of the final state interaction consists of replacing the density operator in the lower band by its average (i.e. neglecting the particle and hole fluctuations.
\begin{equation}
\nonumber
V_{\ssm eff}(x) =  \int dx\,
\biggl[V_{\ssm latt}(x)+\frac{2\pi \bar\rho a_{s}\hbar^{2}}{m}
w_{1i}^{2}(x)
\biggr]
\tilde\psi^{\dag}(x)\tilde\psi(x).
\end{equation}
The effective potential obtained in this way is shown for $^{87}$Rb at $s_{x}=10$
and $\bar\rho=2$  in Fig.~\ref{fig:interactions}(a). The addition of the
Hartree term has two effects. First, it shifts the excited Bloch bands
to slightly higher energies (by $\sim 0.2\, E_{R}$ for the third band)
due to the repulsive interaction with the Mott insulating state. Second,
the bandwidth of the high bands is enhanced by a few percent. This comes
about as the combined effective potential is less confining than the pure
sinusoidal optical lattice, see Fig.~\ref{fig:interactions}(a).

By including the Hartree shift we obtain a new effective Hamiltonian for the particles in the upper bands:
\begin{equation}
H_{eff}=\sum_{n>1,k}e_n(k) b_{nk}^\dagger b_{nk}
\end{equation}
Note that the Bose operators in the upper band are defined using the Wannier states in the effective potential $V_{eff}$:
\begin{equation}
\tilde\psi^{\dag}(x)=\sum_{n>1} \tilde w_{ni}(x) b_{ni}^{\dag}.
\end{equation}

\paragraph{Quasi-particle scattering.} We now turn to the residual interactions which include scattering of particles in the upper band on particles and holes in the lowest band, which is represented by
\begin{equation}
H_{\ssm res}= \sum_{n>1} U_{1n}
 (p_{i}^{\dag}p_{i}-h_{i}^{\dag}h_{i})b_{ni}^{\dag}b_{ni},
\end{equation}
with
\begin{equation}
U_{1n} = \frac{2\pi\bar\rho a_{s}\hbar^{2}}{m}
\int dx\, w_{1,i}^{2}(x)\tilde w_{n,i}^{2}(x).
\end{equation}
Since we are interested in one particular Bloch band, we now suppress the index $n$ from the Bose operators of the upper band and we write $V=U_{1n}$.  We can now write the inter-band interaction using the Bogoliubov quasiparticles of the lowest band defined in \ref{eqn:defbeta}. This gives
\begin{align}
\label{eqn:fullquad}
H_{\ssm res}&=\phantom{+}\frac{V}{N^{2}}
\sum_{k_{1},k_{2},q} s_{k_{1}+q/2,k_{1}}b_{k_{2}-q/2}^{\dag}b_{k_{2}}
\\
\nonumber
&\qquad\qquad \times(\beta_{p,k_{1}+q/2}^{\dag}\beta_{p,k_{1}}-
\beta_{h,-k_{1}-q/2}^{\dag}\beta_{h,-k_{1}}),\\
\nonumber
&\phantom{=}+\frac{V}{N^{2}} \sum_{k_{1},k_{2},q} r_{k_{1}+q/2,k_{1}}b_{k_{2}-q/2}^{\dag}
b_{k_{2}}
\\
\nonumber
&\qquad\qquad \times
(\beta_{p,k_{1}+q/2}^{\dag}\beta_{h,-k_{1}}^{\dag}+
\beta_{p,k_{1}+q/2}\beta_{h,-k_{1}}).
\end{align}
where the interaction coefficients are given by
\begin{align}
s_{p,q}&=u_{p}u_{q}-v_{p}v_{q} \stackrel{J_{1}/U\rightarrow 0}{=} 1, \\
r_{p,q}&=u_{p}v_{q}-u_{q}v_{p} \stackrel{J_{1}/U\rightarrow 0}{=} 0.
\end{align}
The different scattering processes included in the above equations are illustrated in Fig~\ref{fig:interactions}(b)--(e).

We note that processes (b) and (c) are proportional to $u^2$ whereas (d) and (e) scale as $uv$ and are therefore strongly suppressed deep in the Mott phase. Furthermore process (c) vanishes identically because there is only a quasi-hole in the lower band. So, only process (b) contributes significantly to the final state interaction. The contribution of this scattering process to $\Pi(q,w)$ can be computed exactly by resummation of the ladder diagrams shown in Fig.~\ref{fig:interactions}(g).

The resummation reveals the presence of a bound state of a quasi-hole with the high-band particle \cite{Mahan90} at energy of order $V$ below the bottom of the two-particle continuum. The spectral weight of the bound state scales as $(V/J_{n})^{2}\ll1$, where $J_{n}$ is the band-width of the $n$-the Bloch band. We can therefore safely neglect its contribution to $S(q,\omega)$.